# Electrostatic interference control of a high-energy coherent electron beam using a three-element Boersch phase shifter

Pooja Thakkar[1,2,a], Vitaliy A. Guzenko[3], Peng-Han Lu[4], Rafal E. Dunin-Borkowski[4,5], Jan Pieter Abrahams[1,2,6], and Soichiro Tsujino[1,2,b]

[1] Division of Biology and Chemistry, Paul Scherrer Insitut, Forschungsstrasse 111, 5232 Villigen PSI, Switzerland
[2] Swiss Nanoscience Institute, University of Basel, Klingelbergstrasse 82, 4056 Basel, Switzerland
[3] Photon Science Division, Paul Scherrer Institut, Forschungsstrasse 111, 5232 Villigen PSI, Switzerland
[4] Ernst Ruska-Centre for Microscopy and Spectroscopy with Electrons and Peter Grünberg Institute, Forschungszentrum Jülich, Wilhelm-Johnen-Strasse, 52425 Jülich, Germany
[5] Aachen University, Ahornstraße 55, 52074 Aachen, Germany
[6] University of Basel, Klingelbergstrasse 70, 4056 Basel, Switzerland

[a] E-mail: pooja.ummethala@gmail.com
[b] E-mail: soichiro.tsujino@psi.ch

## Abstract

In contrast to static holographic phase shifters, which are restricted to specific electron beam energies and microscope settings, Boersch phase shifters are promising for creating programmable arrays for generating two- and three-dimensional electron beam patterns. We recently demonstrated a three-element Boersch phase shifter device [Thakkar et al., J. Appl. Phys. 128 (2020), 134502], which was fabricated by electron beam lithography and is compatible with up-scaling. However, it suffers from parasitic beam deflection and resulting cross-talk. Here, we report a five-layer phase shifter device, which is based on a metal-insulator-metal-insulator-metal structure (as originally envisioned by Boersch) that reduces cross-talk. We demonstrate a three-element Boersch phase shifter that shows minimal beam deflection of voltage-controlled three-electron-beam interference patterns in a transmission electron microscope operated at 200 keV. The feasibility of using such multi-element phase shifter arrays is discussed.

## 1. INTRODUCTION

Control of the phase of an electron wave front for electron beam shaping is an active area of research for producing arbitrary two- and three-dimensional electron beam patterns.[1,2,3] Holographic phase plates have been used to produce vortex beams[4,5,6,7] and advanced Bessel beams[8,9,10] for spectroscopy and beam propagation control. In order to further exploit this methodology, programmable pixelated phase shifter arrays are needed, which can be used to tune the beam shape to a target sample or for different electron beam energies and microscope settings, as is common in light optics.[11] Such flexibility would have a significant impact on electron wave front engineering applications.[12,13,14,15,16] As a result of the importance of phase distributions for image formation, a programmable device would also be useful for phase contrast imaging,[17] lens-less imaging and holographic image recovery[18,19,20] beyond the use of Zernike phase contrast imaging with static phase plates[21,22,23,24,25,26,27], as well as for the realization of single-pixel electron-beam imaging[28,29,30], which offers possibilities for novel spectroscopic imaging by using an advanced beam analyzer for energy and momentum. There has recently been tremendous progress in the ultrafast modulation of electron wave packets by laser pulses.[31,32,33,34,35,36,37] In contrast, one of the development aims for programmable phase shifter arrays is their integration into state-of-the-art high-resolution electron microscopes without the need for ultrafast modulation of the electron beam or energy filtering.[35]

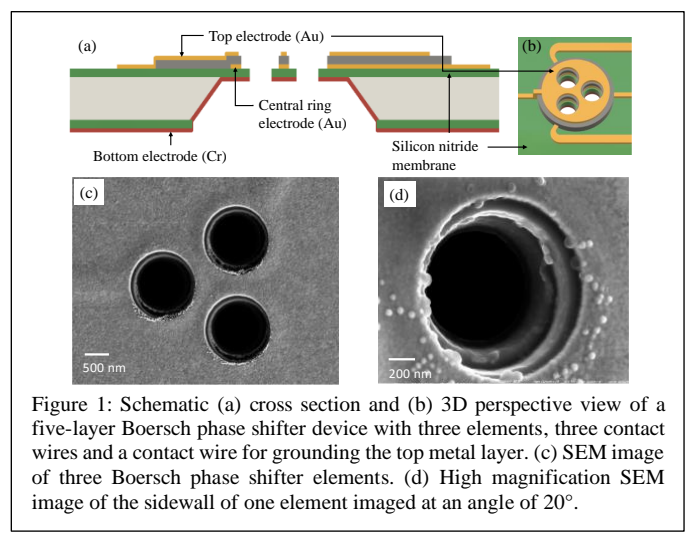

Figure 1: Schematic (a) cross section and (b) 3D perspective view of a five-layer Boersch phase shifter device with three elements, three contact wires and a contact wire for grounding the top metal layer. (c) SEM image of three Boersch phase shifter elements. (d) High magnification SEM image of the sidewall of one element imaged at an angle of 20°.

A Boersch phase shifter device[24,25,26,27,38,39,40] is one of the candidates for a phase shifter technology. It was originally proposed to realize Zernike phase contrast imaging[17] in electron microscopy, and relies on the fact that an electron wave experiences a phase shift when it propagates through an aperture, to which a small electrostatic potential is applied. Although single-element Boersch phase shifters have been tested,[24,25,26,27,38] their fabrication has relied on focused-ion-beam (FIB) milling of stacks of metal and insulator layers. Recently, a four element phase shifter was reported.[14] However, FIB milling is likely to face difficulties for the fabrication of larger numbers of phase shifter arrays, in part because of redeposition.[41]

We recently studied a three-element Boersch phase shifter device, which was fabricated using electron beam lithography in CMOS-compatible processing.[16] We were able to control three-electron-beam interference by applying a few Volts of bias to individual Boersch elements for a 200 keV coherent electron beam.[16] However, as a result of the fact that the simplified three-layer design was based on a metal-insulator-metal structure, the use of unscreened contact wires to the phase shifter elements caused parasitic deflection of the electron beam. This deflection should be eliminated, since cross-talk is difficult to compensate when using a larger array.

The purpose of the present study is to realize a three-element Boersch phase shifter device, in which a top shielding electrode is introduced to prevent parasitic deflection of a far field electron interference pattern. In section II, we describe the design of the device. In section III, we describe the fabrication method of thus design device. In section IV, we describe the result of the experimental characterization of the fabricated device. Finally in section V, we describe the summary and conclusion of the work.

2. DESIGN CONSIDERATION OF BOERSCH PHASE SHIFTER DEVICE

We first investigate the structural parameters for the fabrication of a Boersch phase shifter with a metal-insulator-metal-insulator-metal (MIMIM) five-layer structure. We show electrostatic simulations of a phase shifter element, extend the fabrication procedure from an MIM to an MIMIM structure, and test the device in a high-energy transmission electron microscope (TEM).[42]

Figures 1 (a) and (b) show a schematic cross-section and perspective view, respectively, of a Boersch phase shifter with an MIMIM structure. Figures 1(c) and 1(d) show scanning electron microscopy (SEM) images of a device, in which three metal layers are visible. When it is illuminated by a coherent electron beam, the three transmitted beams form a hexagonal far-field interference pattern. When the three beams have equal phases, the far-field pattern has a bright central spot. When the phase of one of the beams is shifted by applying a voltage, the interference pattern changes and its center becomes dark when the voltage corresponds to a $\pi$-phase shift. The applied phase shift is approximately proportional to the length of the potential distribution in the electron beam direction, but is approximately constant across an aperture.[40] In the MIM structure device reported in Ref. 16, the phase shift was approximately $1.1\pi$ rad/V. However, as a result of the use of unshielded contact wires (CWs) to the phase elements, the beam was also deflected by ~0.35 µrad/V.

In order to study the impact of a top electrode on shielding of the CWs in a MIMIM device, we first simulated the potential distribution around phase shifters without CWs. Figures 2(a-c) show simulations for $dR_{TE} = 0.1$ µm and $V_1 = 1$ V, $V_2 = V_3 = 0$ V, where $dR_{TE}$ is the difference between the inner diameters of the top and ring electrodes (see the inset to Fig. 2(b)) and $V_n$ ($n = 1, 2, 3$) are voltages applied to the $n^{th}$ phase shifter. See also Figure 3 for the detailed definition of the device parameters. We assumed that the dielectric constants for Insulator 1 and Insulator 2 are equal to 7.5 and 3.8, respectively. The thicknesses of both insulators are both equal to 0.2 µm. The Insulator 1 lies at $z = 0$ µm.

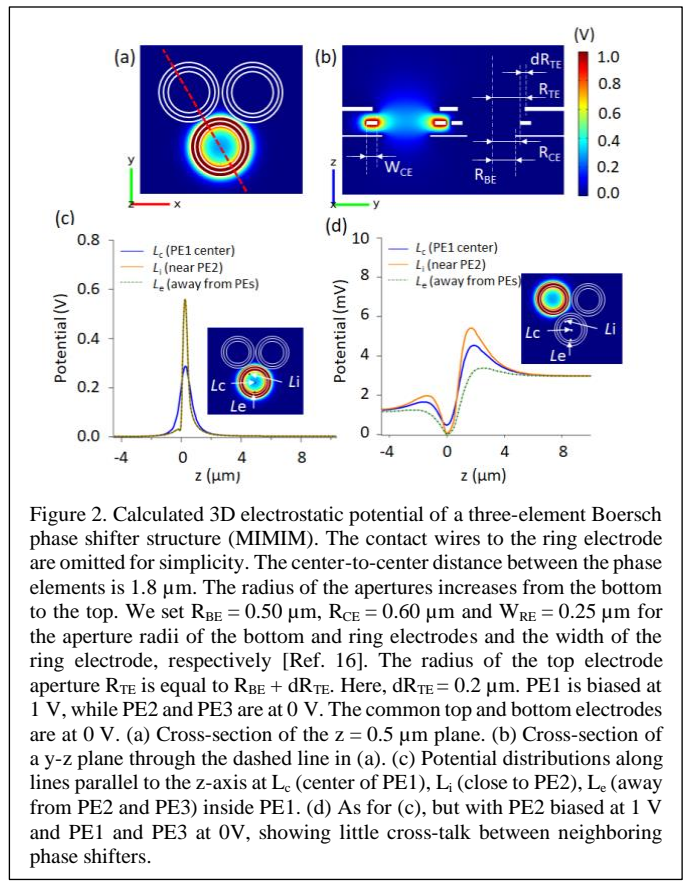

Figure 2. Calculated 3D electrostatic potential of a three-element Boersch phase shifter structure (MIMIM). The contact wires to the ring electrode are omitted for simplicity. The center-to-center distance between the phase elements is 1.8 µm. The radius of the apertures increases from the bottom to the top. We set $R_{BE} = 0.50$ µm, $R_{CE} = 0.60$ µm and $W_{RE} = 0.25$ µm for the aperture radii of the bottom and ring electrodes and the width of the ring electrode, respectively [Ref. 16]. The radius of the top electrode aperture $R_{TE}$ is equal to $R_{BE} + dR_{TE}$. Here, $dR_{TE} = 0.2$ µm. PE1 is biased at 1 V, while PE2 and PE3 are at 0 V. The common top and bottom electrodes are at 0 V. (a) Cross-section of the $z = 0.5$ µm plane. (b) Cross-section of a y-z plane through the dashed line in (a). (c) Potential distributions along lines parallel to the z-axis at $L_c$ (center of PE1), $L_i$ (close to PE2), $L_e$ (away from PE2 and PE3) inside PE1. (d) As for (c), but with PE2 biased at 1 V and PE1 and PE3 at 0V, showing little cross-talk between neighboring phase shifters.

of $\phi_V$ at position $(x, y)$ is given by integration of $U_V(x,y,z)$ along the electron trajectory (the $z$-direction), according to the expression

$$\phi_V(x, y) = \sigma \int_{-\infty}^{\infty} dz\, U_V(x, y, z), \quad (1)$$

where $\sigma = 7.29 \times 10^6$ rad/(Vm) for a 200 keV electron beam.[16,43] Although Eq. (1) is a function of $(x, y)$, $\phi_V$ is approximately constant across an aperture.[40] Figure 4(b) shows that, when compared to the infinite value of $dR_{TE}$ for the MIM case, the small value of $dR_{TE}$ for a screened ring electrode decreases $\phi_V$ by reducing the extension of the potential above the aperture. For the purpose of reliable fabrication, we chose a value of $dR_{TE} = 0.2$ μm, for which the calculated cross-talk is below 2.7 %. We note that in the simulation we assumed that the dielectric constants for Insulator 1 and Insulator 2 are equal to 7.5 and 3.8, respectively. The thicknesses of both insulators are both equal to 0.2 μm. The Insulator 1 lies at z = 0 μm.

In order to evaluate the effectiveness of shielding by the top electrode, we simulated the electrostatic field of a single phase element connected with CWs for $dR_{TE} = 0.2$ μm. We then determined the transverse deflection angle $\eta_s$ from the electric field along the CW: $\eta_s$ is given by the ratio, $u_y/u_z$, where $u_y$ is the transverse velocity and $u_z$ is the longitudinal velocity of the electron when it reaches at a far distance from the phase shifter. For high energy electron beam (equal to 200 keV below), $u_z$ (» $u_y$) is determined by the beam energy. Within the paraxial approximation, $u_y$ is calculated as in Ref. 16, 42, 43,

$$u_y \approx \frac{e}{\gamma m_0 u_z} \int_{-\infty}^{\infty} dz\, F_y(z), \quad (2)$$

where $e$ is the elementary change, $m_0$ is the electron rest mass, and $\gamma m_0$ is the relativistic electron mass ($\gamma = 1.4$ for 200 keV beam energy). $F_y(z)$ is the transverse electric field along the beam trajectory, that is given by the finite element simulation. In the simulation, $u_y$ increases as the calculated propagation distance is extended. In the simulation below (Figure 4), we extended the distance until there is no increase of $u_y$.

Figure 5 shows that the potential of the ring electrode is confined to the phase shifter cavity. As a result, $\eta_s$ for the MIMIM structure is below 0.01 μrad/V, in contrast to 0.28 μrad/V for the MIM structure in theory (0.35 μrad/V in practice).[16]

## 3. FABRICATION OF BOERSCH PHASE SHIFTER DEVICE

Following these considerations, we fabricated a three-element Boersch phase shifter with MIMIM structure with the device parameters as described in Figure 3. In particular, we choose a value for $dR_{TE}$ equal to 0.2 μm.[42] The metal layers are shown in dark red, and Insulator 1 and Insulator 2 are shown in light gray and dark gray, respectively. The top view shows the three phase shifter elements (PE1, PE2 and PE3) in triangular arrangement. The cross-section is through the centre of PE1

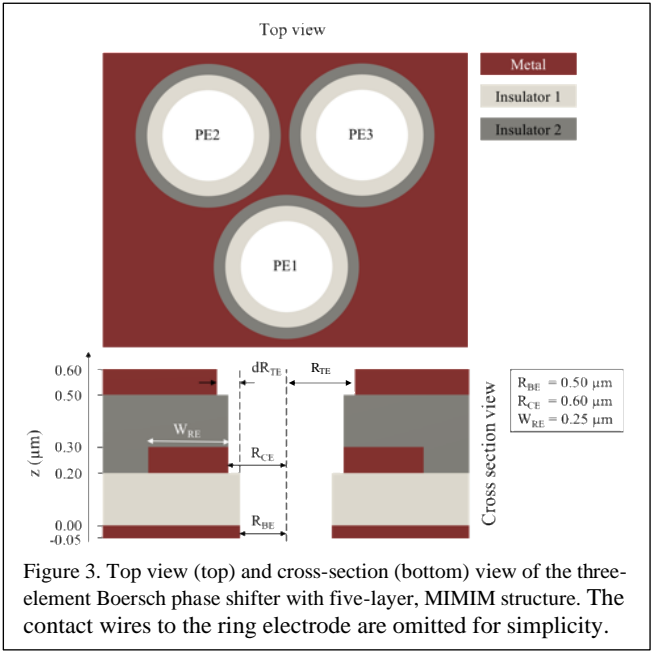

Figure 3. Top view (top) and cross-section (bottom) view of the three-element Boersch phase shifter with five-layer, MIMIM structure. The contact wires to the ring electrode are omitted for simplicity.

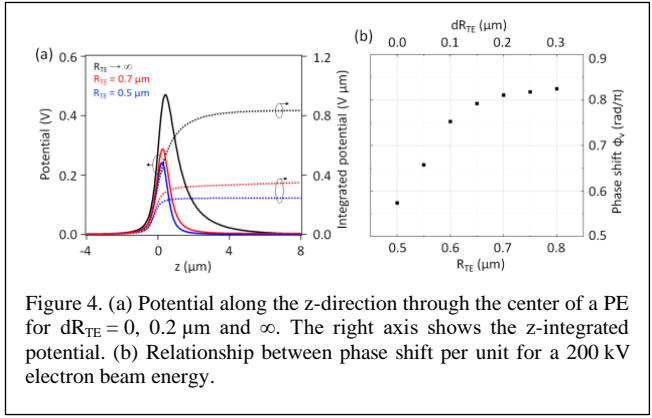

Figure 4. (a) Potential along the z-direction through the center of a PE for $dR_{TE} = 0$, 0.2 μm and ∞. The right axis shows the z-integrated potential. (b) Relationship between phase shift per unit for a 200 kV electron beam energy.

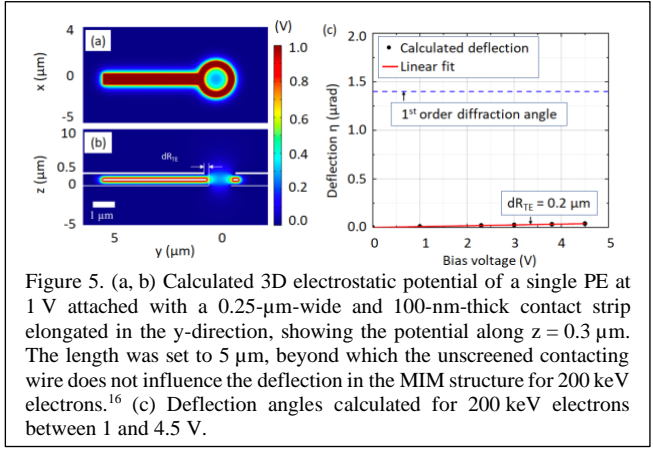

Figure 5. (a, b) Calculated 3D electrostatic potential of a single PE at 1 V attached with a 0.25-μm-wide and 100-nm-thick contact strip elongated in the y-direction, showing the potential along z = 0.3 μm. The length was set to 5 μm, beyond which the unscreened contacting wire does not influence the deflection in the MIM structure for 200 keV electrons.[16] (c) Deflection angles calculated for 200 keV electrons between 1 and 4.5 V.

Figures 4 (a) and (b) show that the potential is confined to the space between the top and bottom electrodes of the phase shifter, as is also apparent from the potential profile $U_V(x,y,z)$ in the beam direction (the $z$-direction) and the phase shift $\phi_V$ for different values of $dR_{TE}$ when the voltage is 1 V. The value parallel to the horizontal axis. The staircase-like structure shows increasing radius of the apertures from bottom to top layers. $R_{BE}$ = the aperture radii at bottom metal and Insulator 1, $R_{CE}$ = the aperture radii at central metal and Insulator 2 layer, being larger than $R_{BE}$ by 100 nm. $W_{RE}$ = 250 nm is the width of the central ring electrode. The radius of apertures in the top most metal layer is equal to $R_{BE} + dR_{TE}$. The dimensions are shown in the inset on bottom right side of the image. The thicknesses of the central and the top metal layers are 0.1 µm. The bottom metal layer thickness is 50 nm. The central ring electrodes are 100-nm-thick and 250 nm wide.

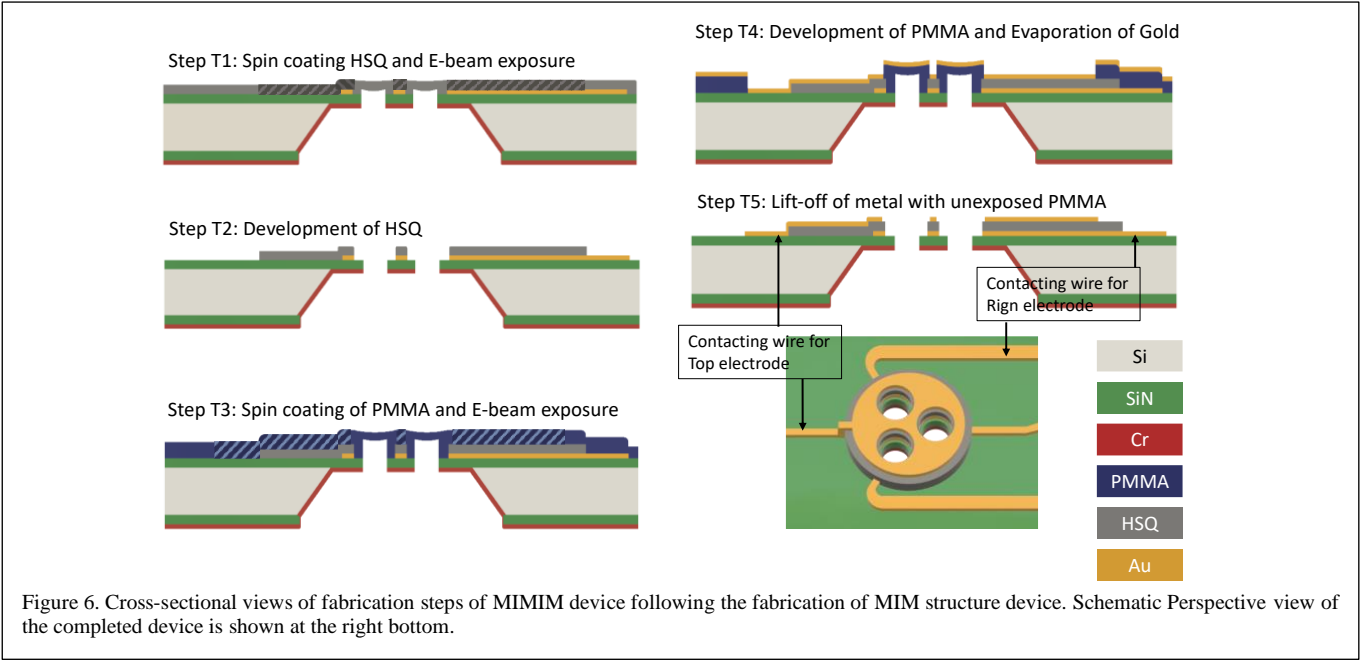

Figure 6. Cross-sectional views of fabrication steps of MIMIM device following the fabrication of MIM structure device. Schematic Perspective view of the completed device is shown at the right bottom.

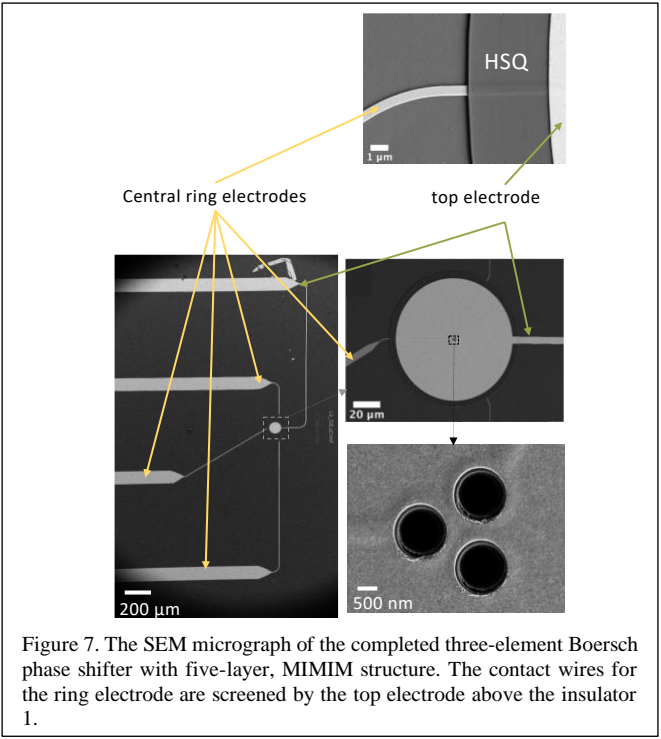

Figure 7. The SEM micrograph of the completed three-element Boersch phase shifter with five-layer, MIMIM structure. The contact wires for the ring electrode are screened by the top electrode above the insulator 1.

The device was fabricated on a low-stress SiN membrane with a thickness of 200 nm, suspended on a 250-µm-thick Si frame. The diameter of the apertures was 1 µm and the separation between their centers was 1.8 µm. Fabrication started with the production of a three-element device with an MIM structure,[16] then follows the steps T1-T5 shown in Figure 6. Insulator 1 was added with cross-linked 200-nm-thick hydrogen silsesquioxane (HSQ), followed by a top-electrode fabricated from a 100-nm-thick Au layer on the top. Insulator 1 was deposited by spin coating, selective electron beam exposure and development. This was followed by electron beam lithography of the Au top electrode and lift-off. During Au evaporation, the resist disk was left on the apertures to protect the sidewalls. For this purpose, we chose a non-zero value of $dR_{TE}$. In addition to Figure 1 (b) and (c), we show in Figure 7 the SEM images of a completed device. CWs for the ring electrodes are screened by the top electrodes above the insulator 1 fabricated with HSQ. CWs are extended to thicker lines that are connected to contacting pads to the far left of the bottom left image (not shown) for mounting the device.

## 4. EXPERIMENTAL CHARACTERIZATION OF FABRICATED BOERSCH PHASE SHIFTER DEVICE

The phase shift performance of the device was tested in a TEM (FEI Tecnai G2 F20) using a beam energy of 200 keV. The sample was mounted in a sample holder (DENSsolutions SH30). The strength of the objective lens was reduced for recording images in far-field diffraction mode.

Figure 8(a) shows a recorded hexagonal interference pattern with the bias voltages to the phase shifters set to zero. Some asymmetry is present due to an offset bias of unknown origin. A symmetrical pattern with a bright central spot was obtained by compensating this offset by applying bias voltages to the phase shifters of $(V_1, V_2, V_3) = (0.8, 0.1, 0)$ V, as shown in Fig. 8(b). Parsitic shift of the whole interference pattern was negligible, as a result of the MIMIM structure shielding the potential of the CWs by the top electrode.

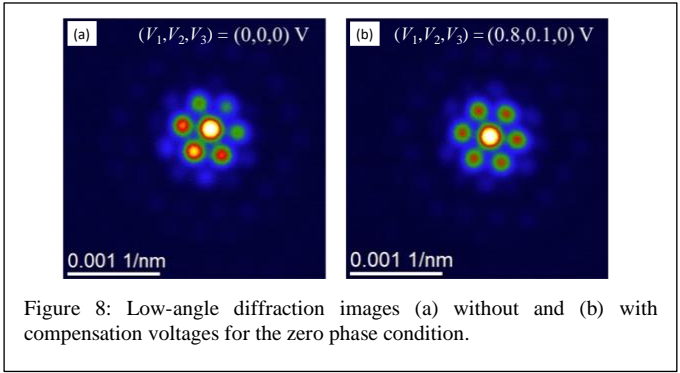

Figure 8: Low-angle diffraction images (a) without and (b) with compensation voltages for the zero phase condition.

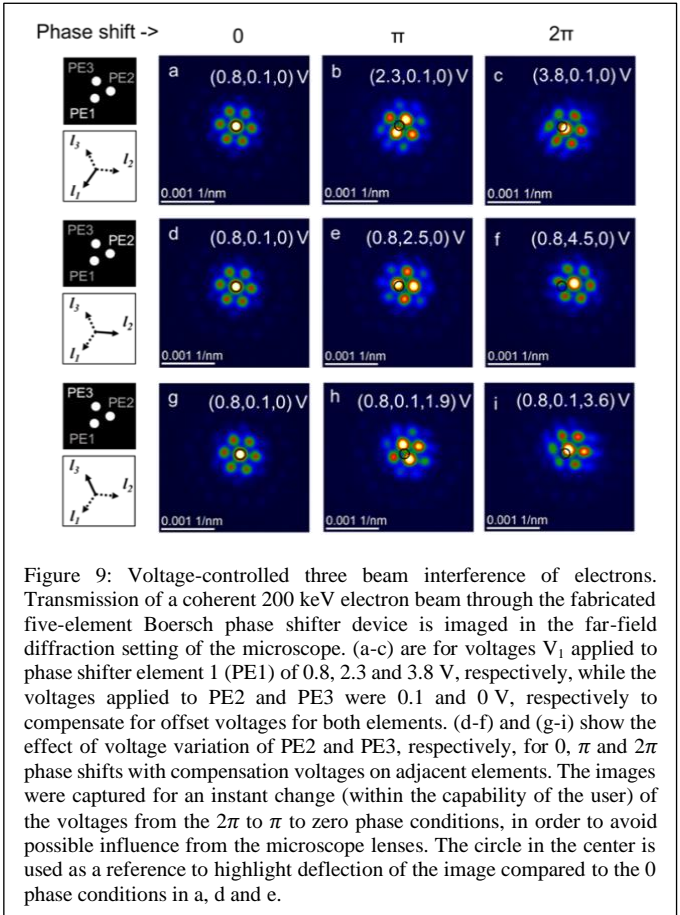

Figure 9: Voltage-controlled three beam interference of electrons. Transmission of a coherent 200 keV electron beam through the fabricated five-element Boersch phase shifter device is imaged in the far-field diffraction setting of the microscope. (a-c) are for voltages $V_1$ applied to phase shifter element 1 (PE1) of 0.8, 2.3 and 3.8 V, respectively, while the voltages applied to PE2 and PE3 were 0.1 and 0 V, respectively to compensate for offset voltages for both elements. (d-f) and (g-i) show the effect of voltage variation of PE2 and PE3, respectively, for 0, $\pi$ and $2\pi$ phase shifts with compensation voltages on adjacent elements. The images were captured for an instant change (within the capability of the user) of the voltages from the $2\pi$ to $\pi$ to zero phase conditions, in order to avoid possible influence from the microscope lenses. The circle in the center is used as a reference to highlight deflection of the image compared to the 0 phase conditions in a, d and e.

Figure 9 shows interference patterns with the individual phase shifters biased for 0, $\pi$, and $2\pi$ phase shifts. For PE1, the observed phase shift was $0.67\pi$ rad/V, which is 0.61 times than the phase shift for the MIM structure, suggesting that the top electrode confined the potential of the ring electrode, as predicted in the simulation shown in Fig. 4(b). The observed phase shift was even smaller for PE2 and PE3, equal to $(0.46\pm0.04)\pi$ and $(0.56\pm0.03)\pi$ rad/V, respectively.

In Figure 9, the position of the central bright spot for a zero phase shift condition is marked by a black circle. When a $\pi$-shift voltage is applied to PE1, the deflection of the interference pattern is 0.12 µrad/V. This is 6.2 times smaller than for the MIM structure with an unshielded CW, demonstrating the effectiveness of the MIMIM structure for reducing parasitic beam deflection. Interestingly, in contrast to the unshielded case, the interference pattern was shifted in a direction unrelated to that of the CW, which is given by the $l_1$ direction from the center of the phase shifters towards the center of PE1 (marked in the insets at the left side of Fig. 9). This result suggests that the observed beam deflection is caused by another mechanism (*e.g.*, the contact on the sample holder). The same small deflection equal to 0.22 and 0.14 µrad/V respectively for PE2 and PE3 in a direction unrelated to the phase shifter position was observed.

The origin of the variation in phase shift with applied voltage between the apertures (of ~20%) is unclear. The phase shift is expected to be proportional to the spatial integration of the potential in the beam direction, suggesting that a difference may arise from a variation in ring-electrode thickness $T_{RE}$, top-electrode aperture size or $dR_{TE}$. High-resolution SEM inspection does not indicate a difference in the value of $T_{RE}$ (nominally 100 nm) of ~20 nm. Referring to the calculated phase shift as a function of $dR_{TE}$ (Fig. 4(b)), the lithographic precision of ~50 nm (in size and alignment) in the value of $dR_{TE}$ of 0.2 µm is too small by more than a factor of 2 to explain the observed phase shift variation of ~20 %. Further research is needed to determine the origin of this variation and to improve the uniformity of phase shifter performance.

## 5. SUMMARY AND DISCUSSION

In summary, we have demonstrated the fabrication of a three-element Boersch phase shifter using a CMOS-compatible method. By adopting an MIMIM structure, we demonstrated minimal parasitic beam deflection and cross-talk, which is important for applications in programmable phase shifter arrays with large numbers of phase elements. The fabricated device shows a $\pi$-shift for a voltage swing of $(2.10\pm0.28)$ V for a 200 keV coherent electron beam. Further research is under way on the design and fabrication of multi-element phase shifter arrays for applications in electron beam shaping, imaging, aberration correction, massively parallel electron beam lithography and the production of arbitrary electron wave fronts.

An important next step is the realization of increased number of PEs. The fabrication of devices with increased numbers of PEs from 3 to ~10 appears to be straightforward, *e.g.*, in circular or hexagonal arrangements for producing vortex beams with the method presented in this work. In contrast to the use of a static holographic mask to produce a vortex beam in a high order diffraction direction, a phase shifter array can be used to produce a vortex beam in the direct beam direction with higher efficiency. Although the offset voltage and phase shift as a function of applied bias have to be known for such applications, their calibration is straightforward, *e.g.*, by determining the voltages that result in a symmetrical interference pattern during measurements. The fabrication of devices with even larger number of PEs *e.g.*, $10^2$-$10^4$ elements promises to be of practical significance for applications such as the correction of spherical aberration.[14] Their production is likely to require a different design of electrical contacts to the individual phase shifters. Solutions adopted for memory devices, such as matrix connections *via* horizontal and vertical wiring with switching elements at each node (corresponding to a phase element in the present case) requires radiation tolerance. Although radiation could result in breakdown for standard MOSFET switches,[44] memristors have recently been studied as potential radiation-tolerant switches.[45,46] Alternatively (in particular for smaller arrays), direct wiring to each of the phase elements is a possibility.[47] In order to minimize space between phase shifters for increasing beam transmission per unit area, a multi-layer design may be needed, together with strain control in the fabrication steps. Although it is desirable to eliminate the phase shift variation and offset voltage, it is feasible to calibrate these effects, *e.g.*, *via* parametric optimization for producing a minimal focal size, since the phase distribution can then be identified as that created by an ideal lens. Nevertheless, we note that such design considerations and fabrication strategy is closely related to the phase shifter functions and performance targeted for the device. For example, as is well known for the optical phase shifter arrays, one should avoid the regular array arrangement to avoid the grating effect.[48] The development of a wiring design strategy and fabrication steps are some of the subjects for future study.

## Acknowledgements

The authors extend their sincere thanks to Prof. Giulio Pozzi (University of Bologna, Italy) for his insight and motivation for this work. They also thank the Laboratory of Micro- and Nanotechnology (LMN) in the Paul Scherrer Insitut (PSI) for support and excellent clean room laboratory infrastructure for device fabrication, as well as Jana Lehmann (LMN, PSI) for help with silicon nitride membrane fabrication. This work was funded by the Swiss Nanoscience Institute Nr. 1505.

## Author contributions

S.T. and J.P.A. conceived the experiment. P.T. conducted the FEM simulation and fabricated the device. V. G. and S.T. supported the simulation and fabrication. P. L. conducted the TEM experiment under the supervision of R. D.-B. and in close communication with P.T. and S.T. P.T. and S.T. analyzed the data and wrote the manuscript. All of the authors read the manuscript and agreed to the contents.